# DEPENDENCE OF SELF-INJECTED BUNCH PARAMETERS ON THE PLASMA DENSITY GRADIENT AND LASER PULSE AMPLITUDE AT LWFA IN A CONICAL PLASMA CHANNEL

*D. S. Bondar[1], V. I. Maslov[1,2], and I. N. Onishchenko[1]*

*[1]National Science Center "Kharkiv Institute of Physics and Technology", Kharkiv, Ukraine*
*[2]Deutsches Elektronen-Synchrotron DESY, Hamburg, Germany*
*E-mail: bondar.ds@yahoo.com*

Laser wakefield acceleration (LWFA) is an advanced method of high gradient acceleration of charged particles with wakefield excited in plasma by laser pulse. A distinctive feature of this method is the ability to create and accelerate so-called self-injected bunches with unique parameters - charge, energy, emittance, and geometric dimensions - without the need for an external injector of the required bunches for their subsequent acceleration to higher energies. Self-injected bunches emerge due to the plasma electrons trapping by the excited wakefield (self-injection phenomenon). For many applications, self-injected bunches can be used directly in relevant experiments. In this paper the dependence of the self-injected bunch parameters on the laser pulse amplitude and longitudinal gradient of plasma density in tapered plasma channel is investigated using numerical simulation with the WarpX code. At the laser amplitude $a_0=E_0(m_{e0}c\omega/e)^{-1}=3.6$ in a conical channel with the radius decreasing from 4.0 $c/\omega_{pe}$ to 2.16 $c/\omega_{pe}$ and with the plasma density increasing linearly from $n_e=2.61\cdot10^{19}$ cm$^{-3}$ at the channel entrance to $3n_e$ at the exit, a self-injected bunch is obtained with the charge 32.1 μC/m, the mean longitudinal momentum 111.9 $m_ec$, the length 5.10 μm, the area 5.3 μm$^2$, the transverse emittance $1.6\cdot10^{-2}$ mm·mrad.

## INTRODUCTION

Laser wakefield acceleration (LWFA) is a promising method for generating and accelerating relativistic electron bunches, and to obtain the high accelerating gradients in plasma [1-5]. Experiments have demonstrated quasi-monoenergetic GeV electron beams from compact setups [6-9].

Formation and dynamics of self-injected bunches was previously studied for dense plasma in [10-12] and in [13-14] for plasma channel obtained by discharge in a capillary used for increasing the acceleration length. Self-injected bunches can be formed in the plasma-dielectric structures at wakefield excitation by electron bunches instead of the laser pulse [15-17].

Joint acceleration by a laser pulse and self-injected bunches as well as hybrid schemes has been considered in [18-21]. A key feature of LWFA is internal injection (so-called self-injection) when background electrons are trapped directly by the wakefield and accelerated without an external injector. The parameters of a self-injected bunch are highly sensitive to the laser and plasma parameters and can be controlled by changing laser driver parameters, using plasma density gradients, laser amplitude, channel profile. In [22-26] shaped plasma density profiles are used to control parameters of self-injected bunches.

Control of the self-injected bunch parameters by laser pulse profile shaping was investigated in [27], and its phase synchronization with the accelerating wakefield in high density plasma in [28-35].

However, in these studies the set of parameters was not considered and the limiting configurations and optimal parameters were not investigated. There are several studies of the generation of high-quality bunches in high density plasma [36], the control of the self-injection threshold by the laser pulse parameters [37, 38].

In this paper some problems are considered that were not investigated before. In particular, it is considered the limit of the inhomogeneity gradient at which the effect of longitudinal momentum increase is offset by strong deformation of the wake bubble and subsequent destroying of the self-injected bunch.

Furthermore, in this paper the dependence of self-injected bunch parameters on laser amplitude is studied and the optimal range within which wake bubble deformation does not lead to the destroying of part of the bunch at the rear wall is to be found. The combined influence of amplitude changes, density gradient and conical channel effect on the shape and size of the wake bubble and, as a consequence, on the self-injected bunch is to be investigated.

The dependence of the self-injected bunch parameters on the laser pulse amplitude $a_0$, channel profile, plasma gradient is investigated, by numerical simulation with the WarpX code [39, 40, 41] in a conical plasma channel for different longitudinally increasing plasma density gradients.

Finding the optimal combination of the density gradient and the laser amplitude is important for obtaining self-injected bunches with needed parameters.

## STATEMENT OF THE PROBLEM

The aim of the study is to determine how the longitudinal plasma density gradient and the laser pulse amplitude influence the parameters of a self-injected electron bunch in a conical plasma channel.

It is needed to show the limits of the parameters within which the deformation of the wake bubble keeps the bunch in the accelerating phase without the destroying of part of the bunch at the bubble rear wall, to find the limiting (too high) density gradient, and to determine the laser amplitude effect for each gradient.

The simulations should be carried out with the fully relativistic particle-in-cell (PIC) code WarpX [39] in 2D3V geometry. A conical plasma channel with a decreasing radius and a longitudinally increasing plasma density will be considered (Fig. 1). Channel radius is 4 $c/\omega_{pe}$ up to z=25 $c/\omega_{pe}$ (cylinder), decreases linearly to 2.16 $c/\omega_{pe}$ at z=85 $c/\omega_{pe}$ (cone). Four configurations will be studied: a homogeneous case $n_e$ and three linear longitudinally increasing gradients with maximum values $2n_e$, $3n_e$ and $4n_e$ at the channel exit (Fig. 1), where $n_e=2.61\cdot10^{19}$ $cm^{-3}$ is the plasma density of the homogeneous section of the channel.

For each configuration the amplitude is chosen as $a_0$=2, 2.4, 2.8, 3.2, 3.6, 4.0. $a_0=E_0(m_ec\omega/e)^{-1}$ is the normalized laser amplitude, where $E_0$ is the peak laser electric field. The laser wavelength is λ=800 nm, $\omega_{pe}$ is defined by $n_{pe}=0.667n_e$, the waist $w_0=3.9$ $c/\omega_{pe}$ and the duration (3σ) 21.7 fs. Both the longitudinal and transverse laser pulse profiles are Gaussian and are truncated at three standard deviations (3σ) from their respective centers. The self-injected bunch parameters are analyzed at time t=238 fs, when the wake bubble head reaches the channel exit on a z=76-90 $c/\omega_{pe}$.

## RESULTS OF SIMULATION

Fig. 1 shows the initial configuration. There are considered cone with homogeneous density $n_e$ or with density gradients rising longitudinally along the cone from $n_e$ up to $2n_e$, $3n_e$ and $4n_e$. Laser pulse and the wake bubble propagate from left to right in the conical channel.

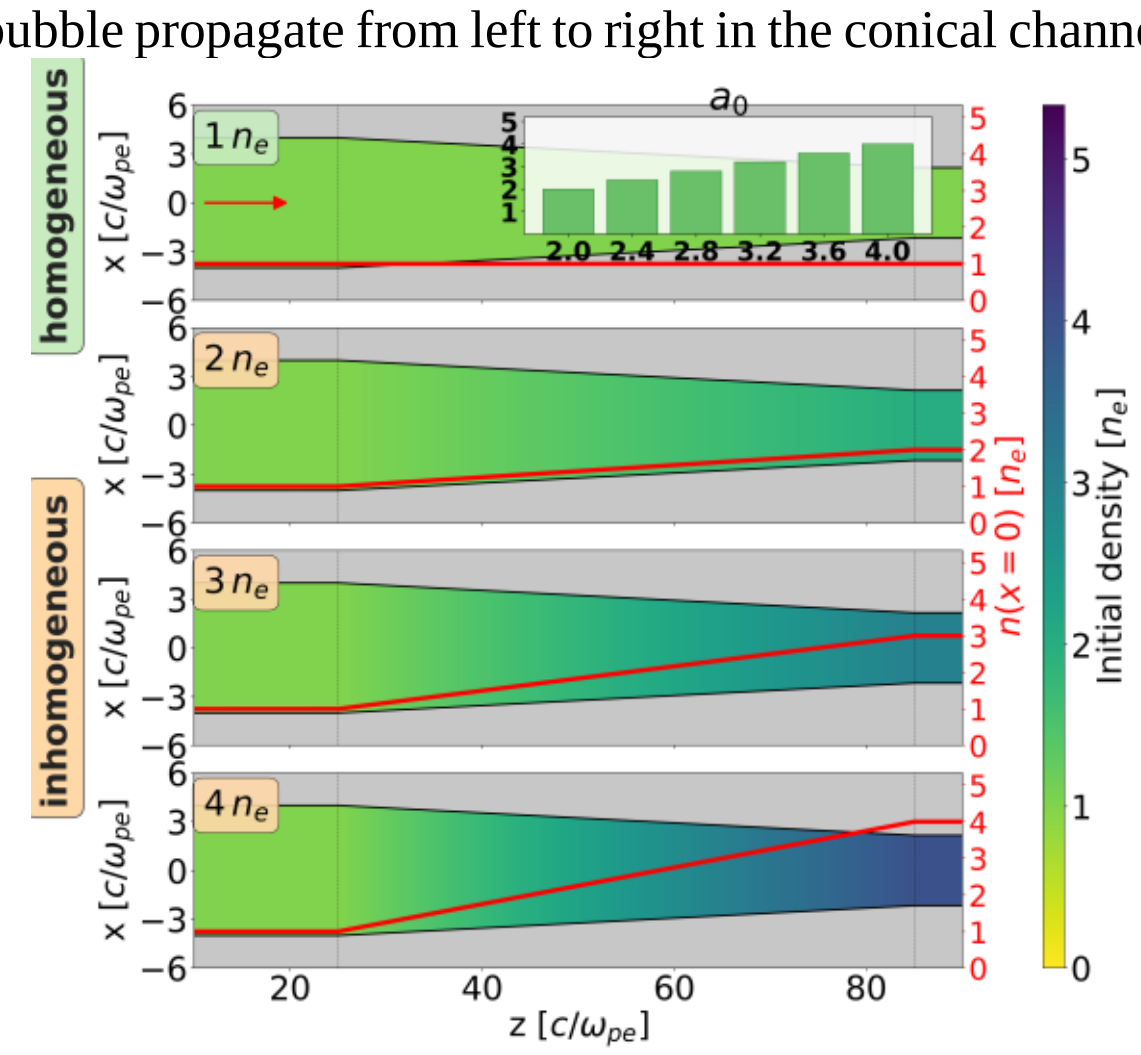

*Fig. 1. Initial electron density (color, in $n_e$) of the converging-cone target in the z-x plane ($c/\omega_{pe}$). Four cases differ in the density ramp along the cone (red line, right axis): homogeneous $n_e$ and inhomogeneous up to $2n_e$, $3n_e$, $4n_e$ at the cone exit (yellow labels). The red arrow marks laser injection. Considered laser amplitudes $a_0$ =2.0.....4.0 (step 0.4). t=238 fs.*

Inside the conical channel laser pulse moves with a radial compression, which increases of laser energy density on the axis of the system that leads to an increase in the efficiency of wake acceleration (see [33]).

It is observed that the maximum laser pulse energy density (Fig. 2) at the system axis is 2.1 times higher (up to $2.82\cdot10^{15}$ J/m³) in the case of the considered conical channel (Fig. 1) compared to the same channel in which the cone is replaced by a cylinder of the constant radius 4 $c/\omega_{pe}$. Both channels are compared under the same conditions, at $a_0$=4, the homogeneous density $n_e$ and the same time t=238 fs, so that only the channel geometry differs. The maximum is reached at the axis at z=86 $c/\omega_{pe}$, near the channel exit ($1.35\cdot10^{15}$ J/m³ in the cylinder and $2.82\cdot10^{15}$ J/m³ in the cone).

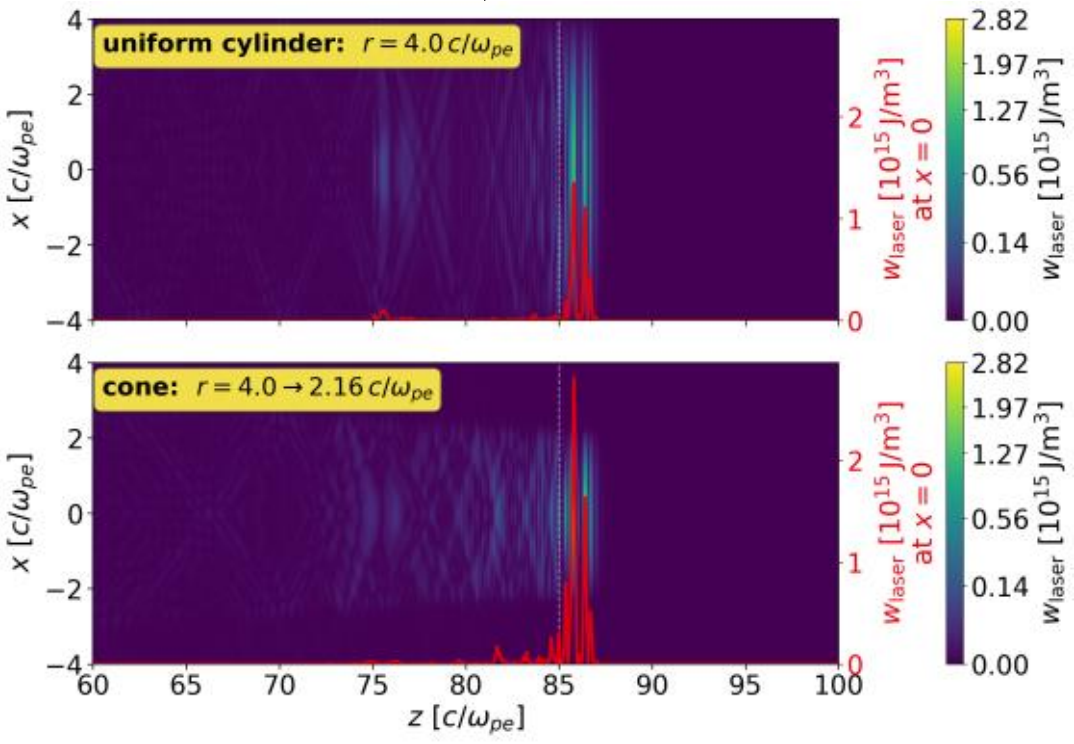

*Fig. 2. Laser pulse energy density $w_{laser}$ in the z-x plane and its value at x=0 for a cylindrical channel (top) and for the cone (bottom). t=238 fs.*

Obtained results of simulation the longitudinal field $E_z$ and electron density $n_e(z,x)$ are shown in Fig. 3 (a, b) for the $a_0$ = 2.0, 2.4, 2.8 (Fig. 3a) and the $a_0$ = 3.2, 3.6, 4.0 (Fig. 3b). Fig. 3a and Fig. 3b present only a fragment near the channel exit, with a length of about the plasma wavelength (z=76-90 $c/\omega_{pe}$).

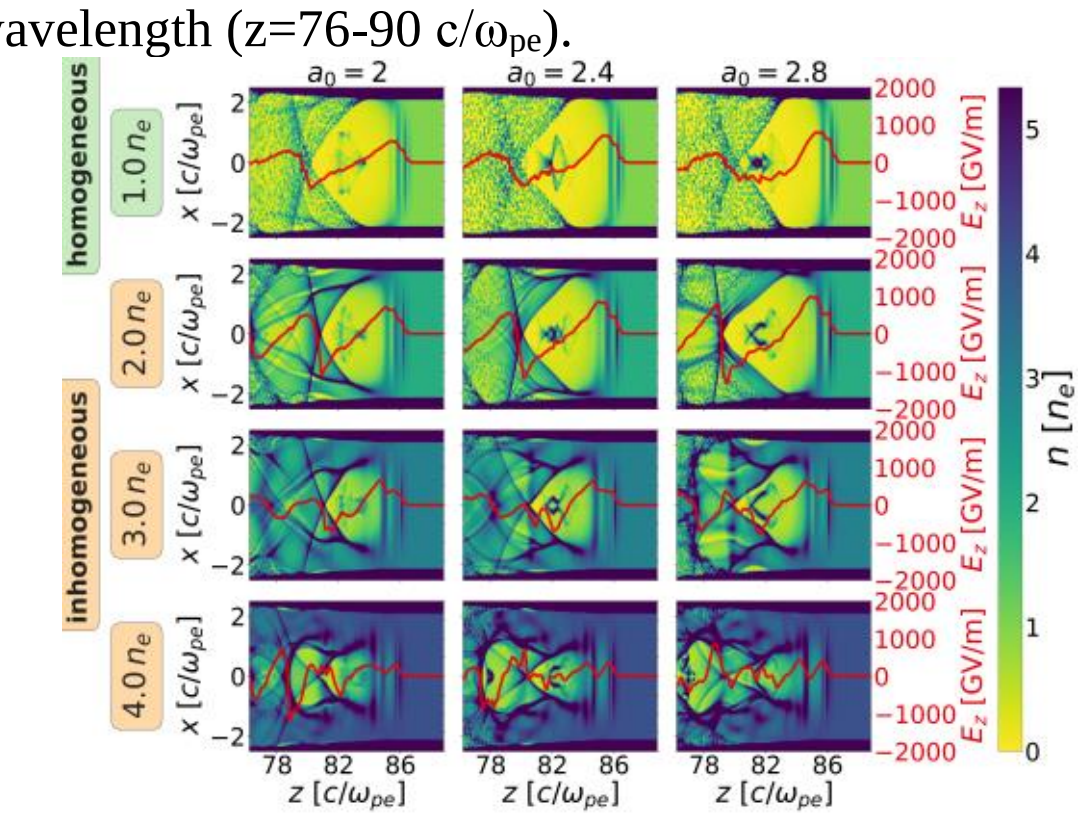

*Fig. 3a. Longitudinal field $E_z$ (red line, GV/m) and electron density $n_e$ (z, x) for laser amplitudes $a_0$ = 2, 2.4, 2.8 (columns) and tapered channel densities: homogeneous $n_e$, inhomogeneous $2n_e$, $3n_e$, $4n_e$ (rows). t=238 fs.*

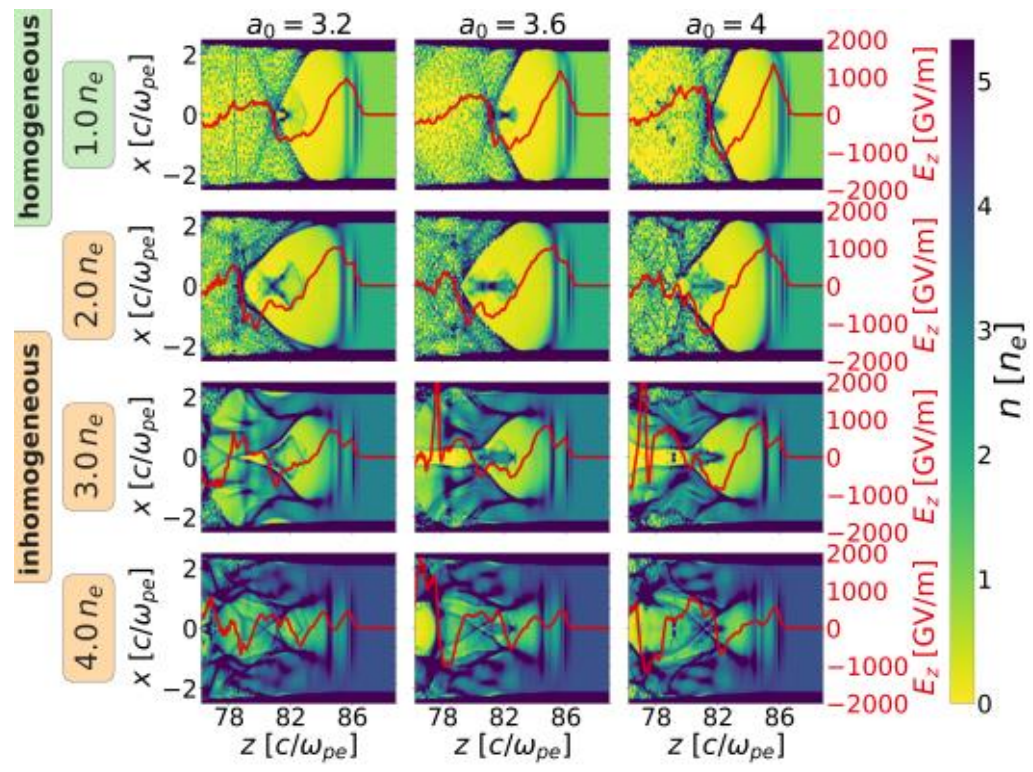

*Fig. 3b. Longitudinal field $E_z$ (red curve, GV/m) and electron density $n_e(z, x)$(colour scale, right) for laser pulse amplitudes $a_0$ = 3.2, 3.6, 4.0 (columns) and tapered channel filled with plasma densities: homogeneous $n_e$, inhomogeneous $2n_e$, $3n_e$, $4n_e$ (rows). t=238 fs.*

The wake bubble is formed behind the laser pulse. To be effectively accelerated the self-injected bunch should be located near the wake bubble tail, where the accelerating field is largest. The accelerating field $E_z$ is about 1300 GV/m (Fig. 3 (a, b), see red curves).

The rough theoretical estimation of bubble size can be taken from [42] $R_b \approx 2\sqrt{a_0}$ $[c/\omega_{pe}]$. For $a_0$=2.0-4.0 it gives bubble radius $R_b$=2.8-4.0 $c/\omega_{pe}$, which already exceeds the channel radius 2.16 $c/\omega_{pe}$ at the cone exit. In the considered geometry the bubble is touched by the channel walls in the homogeneous case.

On Fig. 3 (a, b), in the homogeneous conical channel, the wake bubble is seen to shorten as the amplitude increases. The effect is that, with growing amplitude, the bubble would tend to expand, but the channel walls limit it, so that only the front of the wake bubble is formed and the rear side of the wake bubble is lost. A larger amplitude gives a larger loss and a stronger shortening of the wake bubble along z.

In this case, when a longitudinal gradient of inhomogeneity is used, a decreasing in the plasma wavelength is observed that leads to the decreasing in the wake bubble radius.

In the inhomogeneous cases the bubble is smaller than the channel (the ratios of the channel radius to the bubble radius obtained at the channel exit are 1.13, 1.59 and 2.83 for $2n_e$, $3n_e$ and $4n_e$, they vary only weakly with the amplitude, see Fig. 4).

**Laser pulse amplitude increasing effect.** As seen in Fig. 6, 7 and 8, the bunch length, charge and area grow with $a_0$ up to $a_0$=2.8 and then fall in the homogeneous case and at inhomogeneous $4n_e$. This fall indicates the destroying of part of the bunch. At $2n_e$, $3n_e$ these parameters continue increasing with $a_0$, and the length is largest at $a_0$=3.6. As seen in Fig. 5, $p_z$ grows with $a_0$ in all cases. At $3n_e$ $p_z$ rises from 78 $m_ec$ at $a_0$=2.0 to 126.5 $m_ec$ at $a_0$=4.0. In the homogeneous channel $p_z$ reaches 96.7 $m_ec$ at $a_0$=4.0. At inhomogeneous $4n_e$ the gain of $p_z$ with $a_0$ is smaller than at $3n_e$. The length, charge and area grow with $a_0$ at $2n_e$, $3n_e$ and decrease after $a_0$=2.8 in the homogeneous case and at inhomogeneous $4n_e$, while $p_z$ keeps growing up to $a_0$=4.0.

**Longitudinal density increasing effect.** The longitudinal increase of density shortens the plasma wavelength $\lambda_p$, and the bubble becomes smaller than the channel. At $2n_e$, $3n_e$ the bubble does not touch the channel walls, and the bunch stays in the accelerating phase. This is seen at $a_0$=2.8-3.6. At inhomogeneous $4n_e$ the bubble contracts too strongly and part of the bunch is lost at the rear wall. At $3n_e$ and $a_0$=3.6-4.0 a small plasma channel of length 3.92 $c/\omega_{pe}$ and width 0.67 $c/\omega_{pe}$ forms at the bubble tail (Fig. 3b, 4). At fixed $a_0$ the gradient raises $p_z$ up to $3n_e$. At $a_0$=4.0 $p_z$ reaches 126.5 $m_ec$. At $a_0$=2.0 $p_z$ reaches 78 $m_ec$. At inhomogeneous $4n_e$ the gain in $p_z$ is smaller. A moderate gradient at $2n_e$, $3n_e$ keeps the bunch in the accelerating phase, while inhomogeneous $4n_e$ is excessive and leads to a fall of $p_z$.

The formation of this small channel in the tail region of the wake bubble leads to the deformation of the wakefield bubble structure. Compared with the $2n_e$ case, where no such channel forms, the maximum field $E_z$ at $3n_e$ is lower by 19% at $a_0$=3.6 and by 27% at $a_0$=4.0.

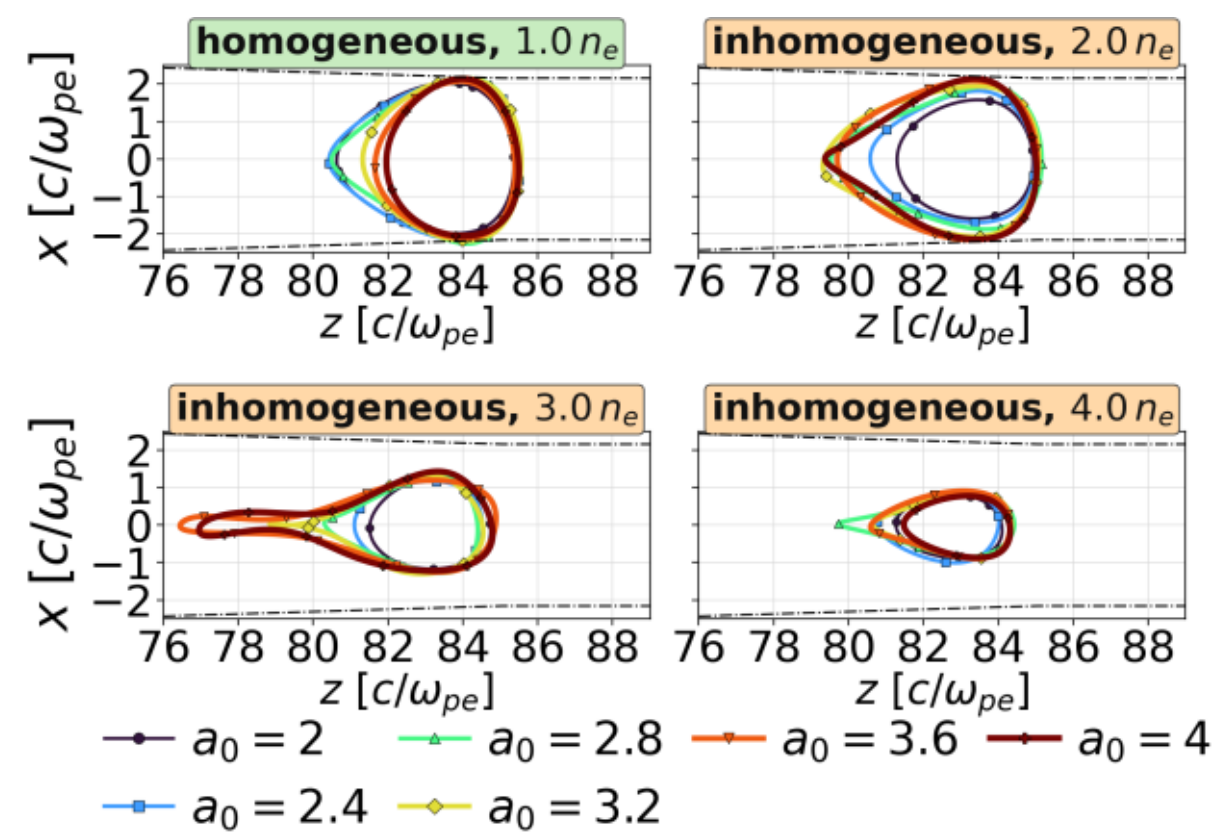

*Fig. 4. Plasma-bubble, one panel per channel density (homogeneous $n_e$, inhomogeneous $2n_e$, $3n_e$, $4n_e$), with one contour per laser amplitude $a_0$ = 2.0-4.0 (step 0.4, line thickness grows with $a_0$). Black dash lines show the tapered channel walls. t=238 fs.*

Fig. 5 shows a graph of longitudinal momentum $p_z$. In the homogeneous case $n_e$ and in the case of inhomogeneous plasma an increase in the average longitudinal momentum $p_z$ is observed with an increase in the laser amplitude $a_0$. In the homogeneous case $n_e$, the maximum value ($a_0$=4.0) of $p_z$ is 96.7 $m_ec$. In the inhomogeneous case $2n_e$ $p_z$ increases to 118.4 $m_ec$. In the inhomogeneous case $3n_e$ $p_z$ increases to 126.5 $m_ec$.

At the same time, in the case $4n_e$, compared to the homogeneous case, $p_z$ increases to 123 $m_ec$. Thus, for $4n_e$, the effect of the $p_z$ increase is even smaller than for $3n_e$.

From Fig. 5 it can be concluded that a case of inhomogeneous plasma with density value of $4n_e$ is too high and leads only to a contraction of the wake bubble, the destroying of part of the self-injected bunch and a decrease in the mean longitudinal momentum $p_z$. This effect is observed for all laser pulse amplitudes.

It is important to note that for a fixed amplitude $a_0$, the use of inhomogeneous plasma with density gradient up to $3n_e$, leads to $p_z$ of 126.5 $m_ec$ at $a_0$=4.0 and 78 $m_ec$ at $a_0$=2.0 (Fig. 5).

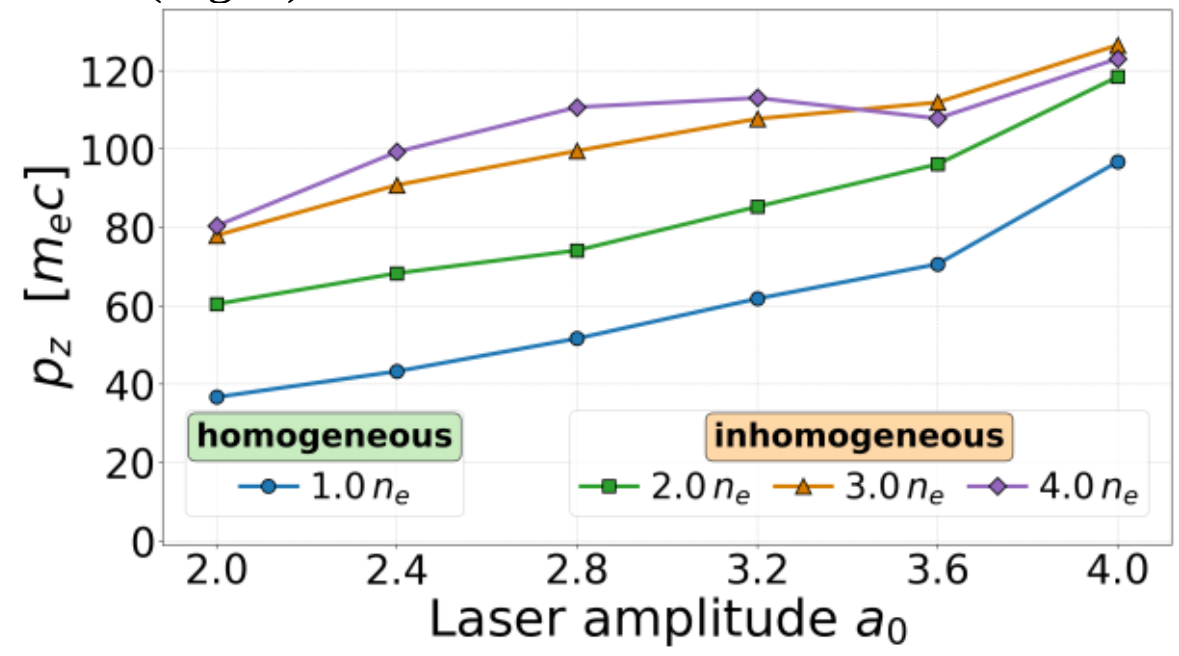

*Fig. 5. Mean longitudinal electron momentum $p_z$ (in $m_ec$) dependence from laser amplitude $a_0$, for the four channel densities (homogeneous $n_e$, inhomogeneous $2n_e$, $3n_e$, $4n_e$). t=238 fs.*

A further increase in the plasma density to $4n_e$ does not lead to an increase in longitudinal momentum $p_z$, due to the previously discussed reason for the significant decrease in the wake bubble size.

In Fig. 6-8 it is presented the main parameters of the bunch length, area and charge. The bunch length (Fig. 6) shows the same threshold behavior as the charge (Fig. 7). In the homogeneous case $n_e$ length grows from 2.98 μm at $a_0$=2.0 to 3.80 μm at $a_0$=2.8-3.2 and then falls to 2.11 μm at $a_0$=4.0.

At $4n_e$ bunch length reaches 3.23 μm at $a_0$=2.8 and falls to 2.02 μm at $a_0$=4.0, which indicates the destroying of part of the bunch at the rear wake bubble wall.

For the moderate gradients $2n_e$, $3n_e$ the length continues increasing with the amplitude growth and reaches 4.93 μm at $2n_e$ and 5.10 μm at $3n_e$ ($a_0$=3.6).

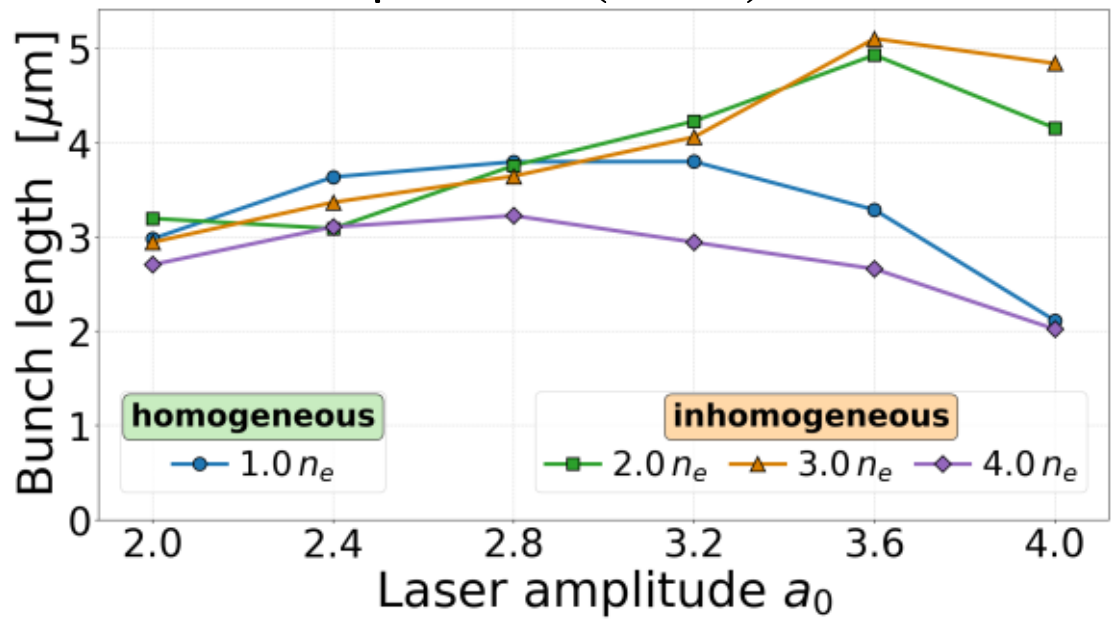

*Fig. 6. Electron bunch length dependence on laser amplitude $a_0$, for the four plasma densities in the channel (homogeneous $n_e$, inhomogeneous $2n_e$, $3n_e$, $4n_e$). t=238 fs.*

Both in the homogeneous case $n_e$, and in the inhomogeneous case with density $4n_e$, a decrease in the charge of the self-injected bunch is observed when reaching high amplitude $a_0$=4.0 (Fig. 7).

In homogeneous case $n_e$ charge changes to 13.5 μC/m at $a_0$=4.0 when initial value at $a_0$=2.0 is the 10.1 μC/m, maximum value at $a_0$=2.8 is the 22.5 μC/m.

At $4n_e$, a fall from 7.8 μC/m to 14.2 μC/m when changing from $a_0$=2.0 to $a_0$=4.0 is observed (maximum value at $a_0$=2.8 is the 20.3 μC/m).

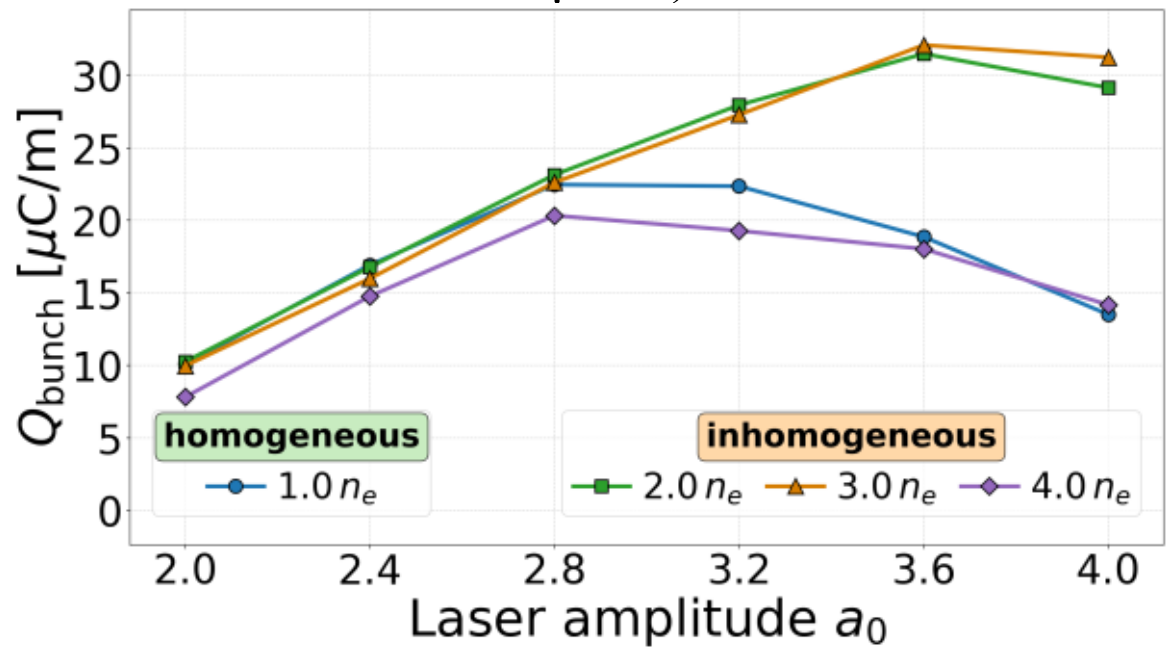

*Fig. 7. Bunch charge $Q_{bunch}$ dependence on normalized laser amplitude $a_0$ for the four plasma densities in the channel (homogeneous $n_e$, inhomogeneous $2n_e$, $3n_e$, $4n_e$). t=238 fs.*

In other cases, for $2n_e$ and $3n_e$, an increase in the charge of the bunch is observed by 2.84 times and 3.13 times respectively when comparing cases $a_0$=4.0 and $a_0$=2.0.

Thus, the use of inhomogeneous plasma leads to an increase in the charge of the self-injected bunch $Q_{bunch}$. This increase shows threshold behavior. In the homogeneous case ($n_e$) or with extreme inhomogeneity ($4n_e$), the charge reaches a maximum and then falls.

In the cases of inhomogeneous increasing densities up to $2n_e$ and $3n_e$, an increase in charge of up to 2.31 times is observed compared to the homogeneous case and the gradient with maximum value $4n_e$.

The dynamic of the self-injected bunch area (Fig. 8) is similar to the charge trend. An increase of the bunch area is observed in the cases $2n_e$, $3n_e$ when amplitude $a_0$ is increasing.

It is observed that in the homogeneous case, as well as in the case of too high inhomogeneous gradient $4n_e$, with an increase in amplitude, the area of the bunch decreases, which indicates the destroying of part of the bunch.

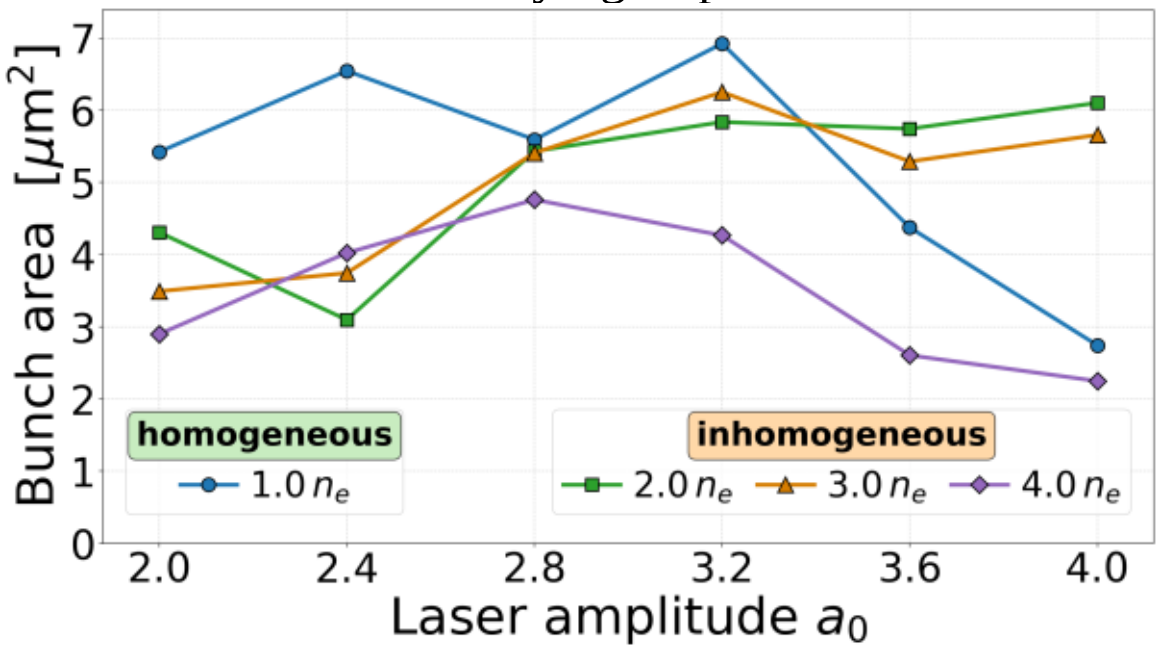

*Fig. 8. Electron bunch area dependence on laser amplitude $a_0$, for the four channel densities (homogeneous $n_e$, inhomogeneous $2n_e$, $3n_e$, $4n_e$). t=238 fs.*

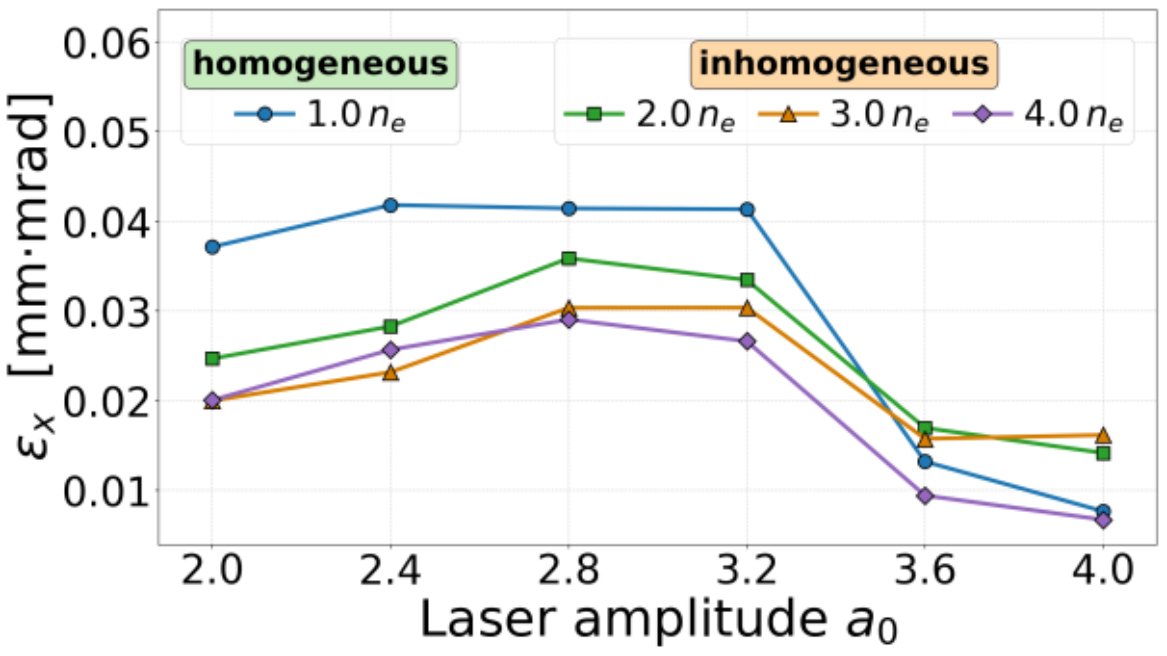

*Fig. 9. Electron bunch transverse emittance $\varepsilon_x$ dependence on laser amplitude $a_0$, for the four channel densities (homogeneous $n_e$, inhomogeneous $2n_e$, $3n_e$, $4n_e$). t=238 fs.*

Fig. 9 shows the emittance graph. The emittance $\varepsilon_x$ (transverse RMS) does not exceed $4.2\cdot10^{-2}$ mm·mrad.

Using an inhomogeneous plasma in a conical plasma channel led to an increase in longitudinal momentum $p_z$ with increasing gradient value until the limiting density is reached ($4n_e$). The effect was observed for all fixed amplitudes. For $a_0$=4.0 $p_z$ reaches 126.5 $m_e c$ at $3n_e$. For $a_0$=2.0 $p_z$ reaches 78 $m_e c$ at $3n_e$.

Thus, it has been demonstrated that when the gradient of the plasma density is fixed, an increase in the value of the longitudinal momentum is observed until the wake bubble contracts so strongly that part of the bunch is destroyed. The obtained results agree with the available experimental data. In [26], an up-ramp plasma density profile increased the mean electron peak energy by more than 50% (from 175±1 MeV to 262±10 MeV), agreed with the growth of $p_z$ obtained here for gradients up to $3n_e$, by 31% at $a_0$=4 and up to a factor of 2.1 at lower amplitudes. Controlled injection in channel-guided accelerators has also been considered and improved by shaping plasma channel [43]. A conical can be formed experimentally in a gas jet by the expansion driven by a laser a preceding laser pulse [44].

## CONCLUSIONS

In this paper the influence of the laser amplitude $a_0$ and the longitudinal plasma density gradient on a self-injected bunch in a conical plasma channel was studied by 2D3V WarpX simulations, for a homogeneous channel $n_e$ and three linear longitudinally increasing plasma densities with maximum values $2n_e$, $3n_e$ and $4n_e$.

Inside the conical channel laser pulse moves with a radial compression, which increases of laser energy density on the axis of the system. At the channel exit the peak laser energy density 2.1 times higher than in a cylindrical channel of constant radius 4 $c/\omega_{pe}$.

The longitudinal density gradient using leads to the mean longitudinal momentum of the self-injected bunch increasing to 126.5 $m_ec$ at $3n_e$ and $a_0$=4.0 compared with 96.7 $m_ec$ in the homogeneous case, while the emittance stays below $4.2\cdot10^{-2}$ mm·mrad. With a gradient increase the wake bubble becomes smaller than the channel with the shortening of the plasma wavelength, so the gradient and the amplitude compensate each other and the bunch is kept in the accelerating phase, its charge exceeding the homogeneous case by up to 2.31 times. The density $4n_e$ is the limit because the bubble contracts too strongly and the momentum reaches only 123 $m_ec$.

Changing the amplitude of the laser pulse $a_0$ affects the longitudinal momentum $p_z$ of the self-injected bunch. With growing $a_0$ at $3n_e$, $p_z$ rises from 78 $m_ec$ at $a_0$=2 to 126.5 $m_ec$ at $a_0$=4, and over the same amplitude range the bunch charge grows 3.13 times. At $2n_e$ and $3n_e$ length and area of the bunch grow with the amplitude, but in the homogeneous channel and at $4n_e$ these parameters pass through a maximum and then fall, which indicates the destroying of part of the bunch. The working range is $a_0$=2.8-3.6 with densities $2n_e$, $3n_e$. At large amplitudes close to $a_0$=4 absorption of the bunch tail and falling of $p_z$ were observed in the homogeneous case.

The optimal obtained bunch ($a_0$=3.6, $3n_e$) has the charge 32.1 μC/m, $p_z$=111.9 $m_ec$, the length 5.10 μm, the area 5.3 μm$^2$ and $\varepsilon_x=1.6\cdot10^{-2}$ mm·mrad.

## ACKNOWLEDGMENTS

The study is supported by the National Research Foundation of Ukraine under the program “Excellent Science in Ukraine” (project # 2023.03/0182).

The work was inspired by discussions with Professor Dr. W. P. Leemans (DESY). We are grateful to Dr. Maxence Thevenet for the useful discussion.

## ЗАЛЕЖНІСТЬ ПАРАМЕТРІВ САМОІНЖЕКТОВАНОГО ЗГУСТКУ ВІД ГРАДІЄНТА ГУСТИНИ ПЛАЗМИ ТА АМПЛІТУДИ ЛАЗЕРНОГО ІМПУЛЬСУ У ЛАЗЕРНОМУ КІЛЬВАТЕРНОМУ ПРИСКОРЕННІ В КОНІЧНОМУ ПЛАЗМОВОМУ КАНАЛІ

*Д. С. Бондар, В. І. Маслов, І. М. Оніщенко*

Лазерне кільватерне прискорення (LWFA) є передовим методом високоградієнтного прискорення заряджених частинок кільватерним полем, збудженим у плазмі лазерним імпульсом. Особливістю цього методу є можливість створювати та прискорювати так звані самоінжектовані згустки з унікальними параметрами - зарядом, енергією, емітансом і геометричними розмірами - без потреби у зовнішньому інжекторі згустків для їх подальшого прискорення до більш вискоих енергій. Самоінжектовані згустки виникають унаслідок захоплення електронів плазми збудженим кільватерним полем (явище самоінжекції). Для багатьох застосувань самоінжектовані згустки можуть використовуватися безпосередньо у відповідних експериментах. У цій роботі досліджено залежність параметрів самоінжектованого згустку від амплітуди лазерного імпульсу, профілю плазмового каналу та поздовжнього градієнта густини плазми за допомогою чисельного моделювання кодом WarpX. За амплітуди лазерного імпульсу $a_0$=3.6 у конічному каналі з радіусом, що спадає від 4.0 $c/\omega_{pe}$ до 2.16 $c/\omega_{pe}$, та з густиною плазми, що лінійно зростає від $n_e$ на вході каналу до $3n_e$ на виході, отримано самоінжектований згусток із зарядом 32.1 мкКл/м, середнім поздовжнім імпульсом 111.9 $m_e c$, довжиною 5.10 мкм, площею 5.3 мкм$^2$ та поперечним емітансом $1.6\cdot10^{-2}$ мм·мрад.